\documentclass[sigconf]{acmart}
\pdfoutput=1
\AtBeginDocument{%
  }

\copyrightyear{2025}
\acmYear{2025}
\setcopyright{rightsretained}
\acmConference[WiSec 2025]{18th ACM Conference on Security and Privacy in
Wireless and Mobile Networks}{June 30-July 3, 2025}{Arlington, VA, USA}
\acmBooktitle{18th ACM Conference on Security and Privacy in Wireless and
Mobile Networks (WiSec 2025), June 30-July}
\acmDOI{10.1145/3734477.3736148}
\acmISBN{979-8-4007-1530-3/2025/06}

\begin{document}

\title[TRIDENT: Privacy in Mobile Propaganda Detection]{TRIDENT - A Three-Tier Privacy-Preserving Propaganda Detection Model in Mobile Networks using Transformers, Adversarial Learning, and Differential Privacy}


\author{Al Nahian Bin Emran\textsuperscript{*}}
\affiliation{%
  \institution{George Mason University}
  \city{Fairfax}
  \state{Virginia}
  \country{USA}}
\email{abinemra@gmu.edu}

\author{Dhiman Goswami\textsuperscript{*}}
\affiliation{%
  \institution{George Mason University}
  \city{Fairfax}
  \state{Virginia}
  \country{USA}}
\email{dgoswam@gmu.edu}

\author{Md Hasan Ullah Sadi\textsuperscript{*}}
\affiliation{%
  \institution{George Mason University}
  \city{Fairfax}
  \state{Virginia}
  \country{USA}}
\email{msadi@gmu.edu}

\author{Sanchari Das}
\affiliation{%
  \institution{George Mason University}
  \city{Fairfax}
  \state{Virginia}
  \country{USA}}
\email{sdas35@gmu.edu}

\thanks{\textsuperscript{*}These authors contributed equally to this work.}

\renewcommand{\shortauthors}{Al Nahian Bin Emran, Dhiman Goswami, Md Hasan Ullah Sadi, \& Sanchari Das}

\begin{abstract}
The proliferation of propaganda on mobile platforms raises critical concerns around detection accuracy and user privacy. To address this, we propose \textbf{TRIDENT}—a \textit{three-tier propaganda detection model implementing transformers, adversarial learning, and differential privacy} which integrates syntactic obfuscation and label perturbation to mitigate privacy leakage while maintaining propaganda detection accuracy. TRIDENT leverages multilingual back-translation to introduce semantic variance, character-level noise, and entity obfuscation for differential privacy enforcement, and combines these techniques into a unified defense mechanism. Using a binary propaganda classification dataset, baseline transformer models (BERT, GPT-2) we achieved F1 scores of $0.89$ and $0.90$. Applying TRIDENT's third-tier defense yields a reduced but effective cumulative F1 of $0.83$, demonstrating strong privacy protection across mobile ML deployments with minimal degradation.
\end{abstract}



\begin{teaserfigure}
  \makebox[\textwidth][c]{%
    \includegraphics[width=0.8\textwidth, height=4cm]{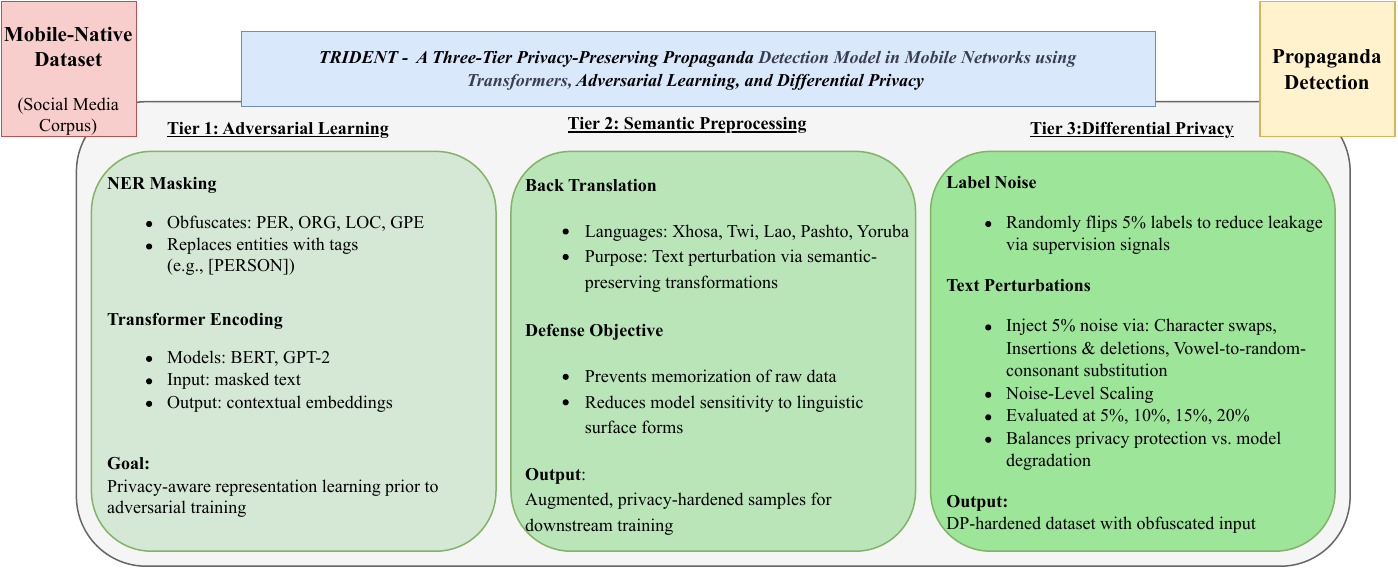}
  }
  \caption{TRIDENT Framework}
  \label{fig:propaganda}
\end{teaserfigure}

\maketitle
\section*{Approach}
Recent research in propaganda detection has increasingly focused on health-related~\cite{sharevski2024debunk} and politically manipulative information~\cite{sharevski2025social}, addressing the spread of misinformation on social media platforms. Transformer-based models such as BERT and GPT-2 have demonstrated strong performance in identifying propaganda content. However, deploying these models in mobile network environments introduces significant privacy risks across the data lifecycle, including data collection, transmission, and inference. To address these concerns, we introduce \textbf{TRIDENT}—a three-tier privacy-preserving framework tailored for mobile-centric propaganda detection. An overview of the TRIDENT workflow is illustrated in Figure ~\ref{fig:propaganda}. TRIDENT integrates (1) named entity masking, (2) multilingual adversarial augmentation, and (3) differential privacy through structured perturbations. Each component is designed to mitigate distinct vectors of privacy leakage, collectively enhancing user data protection while preserving detection accuracy. We evaluate TRIDENT on a binary propaganda classification dataset~\cite{wang2020crossdomain}, consisting of labeled social media posts with a $60/20/20$ train-development-test split. Labels are binary, with `1' representing propaganda and `0' non-propaganda. To ensure fairness and reproducibility, all models were trained with consistent hyperparameters: a batch size of $8$, a maximum token length of $512$, a learning rate of $1 \times 10^{-5}$, weight decay of $0.01$, and $3$ training epochs.

We first established performance baselines using standard fine-tuning of BERT and GPT-2, which achieved test F1 scores of $0.89$ and $0.90$, respectively. We then evaluated each TRIDENT component independently. To assess adversarial robustness, we applied back translation using five linguistically diverse languages: Xhosa, Twi, Lao, Pashto, and Yoruba. This semantic augmentation disrupts shallow memorization in model weights, reducing inference leakage. The resulting models experienced modest performance drops, with F1 scores of $0.85$ (BERT) and $0.86$ (GPT-2). Next, we applied Named Entity Recognition (NER)~\cite{sharma2022named} to anonymize entities categorized as Person (PER), Organization (ORG), Location (LOC), and Geo-Political Entity (GPE). This masking reduces metadata-based re-identification threats. The performance impact was negligible: BERT maintained an F1 of $0.89$, and GPT-2 preserved its original $0.90$ score. For the third tier, we introduced differential privacy (DP) by injecting controlled noise into both labels and text.

Five percent of labels were randomly flipped, and character-level perturbations such as insertions, deletions, and substitutions were applied to 5\% of the training data~\cite{7345591}. This strategy yielded F1 scores of 0.88 for both models. We also experimented with higher noise levels (10\%, 15\%, 20\%), observing a gradual degradation in performance proportional to the perturbation level. Finally, we combined all three privacy mechanisms. With 5\% noise, the full TRIDENT pipeline achieved a robust F1 of $0.83$ for both models—highlighting the trade-off between privacy preservation and detection accuracy. Although performance declined at higher noise rates, both models exhibited similar sensitivity patterns, suggesting architectural agnosticism in the face of privacy constraints. Table~\ref{tab:list3} presents an example of how text is transformed through the combined application of back translation, entity masking, and differential privacy.

\begin{table}[!htp]
\centering
\resizebox{\linewidth}{!}{%
\begin{tabular}{p{3.6cm} | p{7.1cm}}
\hline
\textbf{Techniques} & \textbf{Example (Transformed Data)} \\
\hline
Original (unused in training) & Marilyn from London is calling the RolandMartinShow. \\
\hline
BT, NER, DP (noise = 0.05) & [PERSlN] calmle dthN Roland MJrtin Show from [GPE]. \\
BT, NER, DP (noise = 0.10) & [PERSON] cRlled the Roan djMartin Sohw from [GD]. \\
BT, NER, DP (noise = 0.15) & [EPRSONo]callc ftvhe RKgland Martin Shcw from [GPE]. \\
BT, NER, DP (noise = 0.20) & [PRrSlN] ucalled the Rolwnd MTrtni Show from v[GP]zz. \\
\hline
\end{tabular}
} 
\caption{Text Transformations With Increasing Privacy Noise}
\label{tab:list3}
\vspace{-8mm}
\end{table}

Our experiment highlight a fundamental trade-off between model accuracy and privacy preservation. Adversarial augmentation via multilingual back-translation proved effective in reducing inference-based privacy risks, offering protection against model memorization without severely impacting performance. NER contributed to protecting metadata by anonymizing identifiable tokens, directly addressing concerns related to storage and re-identification threats. Notably, NER produced negligible performance degradation, maintaining high F1 scores across both transformer models. DP, which targeted vulnerabilities during data collection and preprocessing, provided stronger protection by perturbing labels and textual content. While effective in increasing privacy guarantees, this technique introduced measurable accuracy losses—particularly as noise levels increased. Table~\ref{tab:list3} illustrates the progressive degradation in sample quality and model performance as perturbation intensity scales from $5\%$ to $20\%$.

\begin{table}[!htp]
\centering
\scalebox{0.685}{
\begin{tabular}{c|ccc}
\hline
\textbf{Base Model} & \textbf{Procedure} & \textbf{Dev F1} & \textbf{Test F1}\\
\hline
& Direct Fine-tune & 0.89 & \bf 0.89\\
\cline{2-4}
& Adversarial Defense (Back Translation) & 0.84 & 0.85\\
& NER Masking & 0.87 & 0.89\\
& DP (noise = 0.05) & 0.83 & 0.88\\
& DP (noise = 0.10) & 0.77 & 0.86\\
BERT & DP (noise = 0.15) & 0.71 & 0.81\\
& DP (noise = 0.20) & 0.67 & 0.75\\
\cline{2-4}
& BT, NER, DP (noise = 0.05) & 0.74 & \bf 0.83\\
& BT, NER, DP (noise = 0.10) & 0.69 & 0.80\\
& BT, NER, DP (noise = 0.15) & 0.64 & 0.75\\
& BT, NER, DP (noise = 0.20) & 0.62 & 0.64\\
\hline
& Direct Fine-tune &  0.90 & \bf 0.90 \\
\cline{2-4}
& Adversarial Defense (Back Translation) & 0.83 & 0.86\\
& NER Masking &  0.87 & 0.90 \\
& DP (noise = 0.05) &  0.84 & 0.88 \\
& DP (noise = 0.10) &  0.78 & 0.86 \\
GPT-2 & DP (noise = 0.15) &  0.73 & 0.81 \\
& DP (noise = 0.20) &  0.68 & 0.78 \\
\cline{2-4}
& BT, NER, DP (noise = 0.05) &  0.74 & \bf 0.83 \\
& BT, NER, DP (noise = 0.10) &  0.71 & 0.80 \\
& BT, NER, DP (noise = 0.15) &  0.66 & 0.76 \\
& BT, NER, DP (noise = 0.20) &  0.64 & 0.68 \\
\hline
\end{tabular}
}
\caption{Experimental Results of Propaganda Detection}
\label{tab:result}
\vspace{-8mm}
\end{table}

When all three privacy-preserving components were applied simultaneously, the TRIDENT framework achieved robust protection across the model pipeline from raw data input to inference—while maintaining an acceptable reduction in performance. With a $5\%$ noise level, both BERT and GPT-2 achieved a final F1 score of $0.83$, underscoring the framework's capability to preserve model utility under compound defenses. Interestingly, both encoder-based (BERT) and decoder-based (GPT-2) models exhibited similar sensitivity trends when exposed to noisy or augmented training data. This suggests that TRIDENT's privacy-preserving techniques generalize well across different model architectures, which is particularly valuable for varied mobile deployment scenarios. These findings emphasize the practical applicability of privacy-enhancing mechanisms within mobile machine learning workflows. Each technique addresses specific threats within the data lifecycle ranging from preprocessing to inference highlighting the importance of adopting a layered defense strategy. A summary of model performance across different privacy settings is presented in Table~\ref{tab:result}.

\bibliographystyle{ACM-Reference-Format}
\bibliography{sample-base}

\end{document}